\documentclass{elsart}
\usepackage{epsfig}
\begin{document}

\begin{frontmatter}
\title{Noise-free Stochastic Resonance in Simple Chaotic Systems}

\author{Sitabhra Sinha}\footnote{E-mail: sitabhra@physics.iisc.ernet.in}

\address{Department of Physics,
Indian Institute of Science,
Bangalore - 560 012, India\\
and\\
Condensed Matter Theory Unit, Jawaharlal Nehru Centre for Advanced
Scientific Research, Jakkur, Bangalore - 560 064, India.}

\begin{abstract}
The phenomenon of Stochastic Resonance (SR) is reported in
a completely noise-free situation, with the role of thermal noise
being taken by low-dimensional chaos. A one-dimensional,
piecewise linear map and
a pair of coupled excitatory-inhibitory neurons are the systems
used for the investigation. Both systems show a transition from
symmetry-broken to symmetric chaos on varying a system parameter.
In the latter state, the systems switch between the formerly
disjoint attractors due to the inherent chaotic dynamics. This
switching rate is found to ``resonate'' with the frequency
of an externally applied periodic perturbation (either parametric
or additive). The existence of a resonance in the response of the
system is characterized in terms of the residence-time distributions.
The results are an unambiguous indicator of the presence of
SR-like behavior in these systems. Analytical investigations
supporting the observations are also presented.
The results have
implications in the area of information processing in biological
systems.

\vspace{0.25cm}
PACS nos.: 05.40.+j, 05.45.+b
\end{abstract}

\end{frontmatter}

\section{Introduction}
``Stochastic Resonance'' (SR) is a recently observed cooperative phenomena 
in nonlinear
systems, where the ambient noise helps in amplifying a sub threshold
signal (which
would have been otherwise undetected) when the signal frequency is close
to a critical value \cite{Ben81} (see \cite{Gam98} for a recent review).
A simple scenario for observing such a
phenomena is a heavily damped bistable dynamical system (e.g., a potential
well with two minima) subjected to an external periodic signal. As a 
result, each of the minima are alternately raised and lowered in the
course of one complete cycle. If the
amplitude of the forcing is less than the barrier height between the wells,
the system cannot switch between the two states. However, the introduction
of noise can give rise to such switching. As the noise level is gradually
increased, the stochastic switchings will approach a degree of synchronization
with the periodic signal until the noise is so high that the bistable
structure is destroyed, thereby overwhelming the signal.
So, SR can be said to occur because of noise-induced
hopping between multiple stable states of a system, locking on to an
externally imposed periodic signal.

The characteristic signature of
SR is the non-monotonic nature of the {\em Signal-to-Noise Ratio} (SNR)
as a function of the external noise intensity. 
A theoretical understanding of this phenomena in 
bistable systems, subject to both periodic and random forcing,
has been obtained based on the rate equation
approach \cite{Mcn89}. As the output of a
chaotic process is indistinguishable
from that of a noisy system, the question of whether a similar process
occurs in the former case has long been debated. In fact, Benzi {\it et al}
\cite{Ben81} indicated that the Lorenz system of equations, a well-known
paradigm of chaotic behavior might be
showing SR. Later studies \cite{Nic93} 
in both discrete- and continuous-time systems seemed to
support this view. However, it is difficult to guarantee that the response
behavior is due to ``resonance'' and not due to ``forcing''.
In the latter case, the periodic perturbation is of so large an
amplitude, that the system is forced to follow the driving frequency
of the periodic forcing. The ambiguity is partly because the
SNR is a monotonically decreasing function of
the forcing frequency and cannot be used to distinguish
between resonance and forcing. 

Signature of SR can also be observed in the {\em residence time distribution}. 
In the presence of a periodic modulation, the distribution shows
a number of peaks superposed on an exponential background. However,
this is observed both in the case of resonance as well as
forcing. The ambiguity is, therefore, present in 
theoretical \cite{Cri94} and experimental \cite{Rei97} studies
of noise-free SR, where regular and chaotic phases take the role
of the two stable states in conventional SR. Although the distribution
of the lengths of the chaotic interval shows a multi-peaked structure,
this by itself is not sufficient to ensure that the enhanced response
is not due to ``forcing''.
In the present work this problem is avoided by measuring the response of
the system in terms of the peaks in the normalized distribution of 
residence times \cite{Gam95}. For SR, the strength of the peaks 
shows non-monotonicity with the
variation of both noise intensity and signal frequency. 

In this paper we present two simple models for studying stochastic resonance
where the role of noise is played by the chaos generated through the
inherent dynamics of the system. 
In Section 2, the first
model for studying deterministic SR is introduced. It is
a 1-dimensional piecewise linear map with uniform slope
throughout. The numerical
observation of resonance in computer simulations for
parametric perturbation is described and
a theoretical analysis of these
observations is given. Additive perturbations also give rise 
to similar resonance behavior.
In Section 3, we consider the second model, an
excitatory-inhibitory neural pair. It is additively perturbed
with a very low amplitude signal and the system response is observed, 
for which numerical and theoretical
results are given. We conclude with a discussion on the 
implication of such resonance phenomena for biological systems. 

\section{The 1-dimensional map model}
Recently, SR has been studied in 1-D maps with two well-defined states 
(but not necessarily stable) with switching between them aided
by either additive or multiplicative external noise \cite{Gad97}.
However, dynamical contact of two chaotic 1-D maps can also induce
rhythmic hopping between the two domains of the system \cite{Sek96}.
We now show how the
chaotic dynamics of a system can itself be used for resonant switching
between two states, without introducing any external noise.

The model chosen here is a piecewise linear anti-symmetric map,
henceforth referred to as the Discontinuous
Anti-symmetric Tent (DAT) map \cite{Sin98a}, defined in the interval [$-1,
1$]:
\begin{equation}
x_{n+1} = {\rm F}(x_n) =\left\{
\begin{array}{ll}
  1 + a(0.5-x_n),  & {\rm if}~~x_n \geq 0.5\\
  1 - a(0.5-x_n),  & {\rm if}~~0 < x_n < 0.5\\
 -1 + a(0.5+x_n), & {\rm if}~~-0.5 < x_n < 0\\
 -1 - a(0.5+x_n),  & {\rm if}~~x_n \leq -0.5.
\end{array}
\right.
\end{equation}
The map has a discontinuity at $x = 0$. The behavior of the system is
controlled by the parameter $a$ $(0<a<4)$. 
The map has a symmetrical pair of fixed points $x^*_{1,2}= \pm {\frac
{1+a/2}{1+a}}$
which are stable for $0<a<1$ and unstable for $a>1$. Another pair of
unstable fixed points, $x^*_{3,4} = \pm {\frac{1-a/2}{1-a}}$ come into
existence for $a>2$. 
Onset of
chaos occurs at $a=1$. The chaos is symmetry-broken,
i.e., the trajectory is restricted
to either of the two sub-intervals R:($0, 1$] and L:[$-1,
0$), depending on
initial condition. Symmetry is restored at $a=2$.
It is to be noted that as $a \rightarrow 2$ from
above, $x^*_{3,4}$ both collide at $x=0$ causing an interior crisis,
which leads to symmetry-breaking of the chaotic attractor.
The Lyapunov exponent
of the map is a simple monotonic function of the parameter $a$. 

To observe SR, the value
of $a$ was kept close to 2, and then modulated sinusoidally with amplitude
$\delta$ and frequency $\omega$, i.e.,
\begin{equation}
a_{n+1} = \left \{
\begin{array}{ll}
a_{0} ~+~ \delta~ {\rm sin}(2 \pi \omega n), & {\rm if}~~x \in {\rm R}\\
a_{0} ~-~ \delta~ {\rm sin}(2 \pi \omega n), & {\rm if}~~x \in {\rm L}.
\end{array}
\right.
\end{equation}
We refer to this henceforth as multiplicative or parametric perturbation,
to distinguish it from additive perturbation (discussed later).

\begin{figure}
\begin{center}
\leavevmode
\epsfig{file=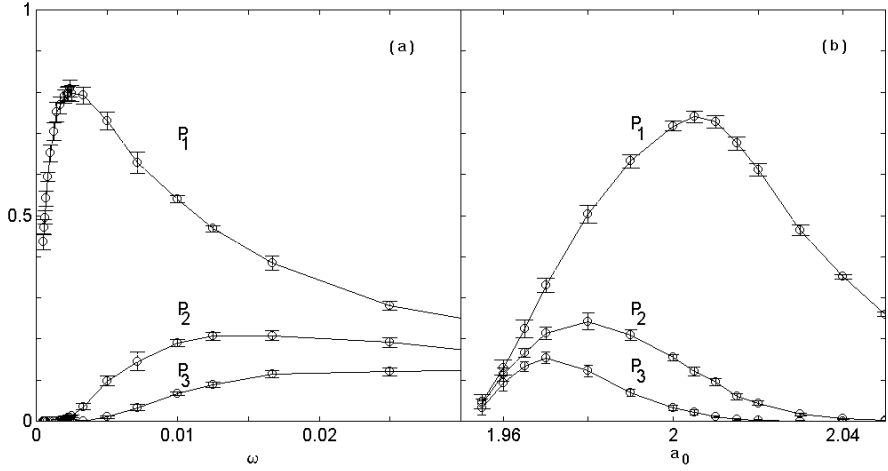, height=2.75in}
\end{center}
\caption{(a) $P_n$ ($n$ = 1, 2, 3) versus $\omega$ for $a_0 = 2.01$ and
$\delta = 0.05$, (b) $P_n$ ($n$ = 1, 2, 3) versus $a_0$ for $\omega =
1/400$ and $\delta = 0.05$. The circles represent the average value of
$P_n$ for 18 different initial values of $x$, the bars representing the
standard deviation.}
\label{fig204}
\end{figure}

The system immediately offers an analogy to the classical bistable well
scenario of SR. The sub intervals L and R
correspond to the two wells between
which the system hops to and fro, aided by the inherent noise (chaos) and
the external periodic forcing. 
The response of the system is measured in terms of
the normalized distribution
of residence times, $N(n)$ \cite{Gam95}. This distribution shows a series
of peaks centered at $n_j = (j - {\frac{1}{2}}) n_0$, i.e., odd-integral
multiples of the forcing period, $n_{0} = {\frac{1}{\omega}}$. 
The strength of the $j$-th peak
\begin{equation}
P_j = \int_{n_j - \alpha n_0}^{n_j
+ \alpha n_0} N(n) dn ~~~(0<\alpha<0.25),
\label{eq303}
\end{equation}
is obtained  at different values of $\omega$, keeping $a_0$
fixed for $j$=1,2 and 3. To maximize sensitivity, $\alpha$
was taken to be $0.25$.
For $a_0=2.01$ and $\delta=0.05$, the response of the system showed a
non-monotonic behavior as $\omega$ was varied, with $P_1$ peaking
at $\omega_1 \sim 1/400$, a value dependent upon $a_0$ $-$ a clear
signature of SR-type phenomenon. $P_2$ and $P_3$ also showed non-monotonic
behavior, peaking roughly at odd-integral multiples of $\omega_1$
(Fig. \ref{fig204}~(a)). 
Similar observations of $P_j$ were also done by varying $a_0$,
keeping $\omega$ fixed. Fig. \ref{fig204}~(b) shows the results of
simulations for $\omega = 1/400$ and
$\delta=0.05$. Here also a non-monotonicity was observed
for $P_1$,$P_2$ and $P_3$.
The broadness of the response curve and the magnitude of the
peak-strengths
are a function of the perturbation magnitude, $\delta$.

Analytical calculations were done to obtain the average residence-time
at any one of the sub intervals. This gives the dominant time-scale of the
intrinsic dynamics. Mapping the system dynamics to an approximately
first-order Markov process, the mean residence time is obtained as \cite{Sin98a}
\begin{equation}
< n > = {\frac {-1}{{\log}({\frac{1-{\epsilon}/2
-{\epsilon}^2/4}{1-{\epsilon}^2/4}}) }} \simeq
{\frac{-1}{\log(1-\epsilon/2)}},
\label{avgeq}
\end{equation}
where, $\epsilon = a_0 - 2$.
So, for $a_0=2.01$, $< n > \simeq 200$. 
This predicts that a peak in the response should be observed at a frequency
${\frac{1}{2 < n >}} \simeq 1/400$, which agrees with the simulation results.

The mean time spent by the trajectory in any one of the 
sub-intervals (L or R) can also be
calculated exactly for piecewise linear maps \cite{Eve87}. 
From the geometry of the DAT map, the total
fraction of R which escapes to L after $n$ iterations is found to be
$l_n = {\frac{2^n \epsilon}{2 (2 + \epsilon)^n}}$ \cite{Sin98a}.
This is just the probability that the trajectory spends a period
of $n$ iterations in R before escaping to L ($\sum_{j=1}^{\infty}
l_j = 1$). So the average lifetime of a trajectory
in R is
\begin{equation}
< n > = \sum_{j=1}^{\infty} (j-1) l_j = {\frac{2}{\epsilon}}.
\end{equation}
Note that, as $\epsilon \rightarrow 0$, Eqn. (\ref{avgeq}) becomes
identical to the above expression.
For $a_0$ = 2.01, $< n >$ = 200, in good agreement with the result
obtained using the approximate Markov partitioning (which ensures the
validity of the approximation). By symmetry
of the map, identical results will be obtained if we consider
the trajectory switching from L to R.

Similar study was also conducted with additive perturbation for
the above map. In this case the dynamical system is defined as
$x_{n+1}={\bf \rm F}(x_{n}) ~+~ \delta {\rm sin}(2 \pi \omega n)$.
The simulation results showed non-monotonic behavior for the response, as
either $\omega$ or $a_{0}$ was varied, keeping the other constant, but
this was less marked than in the case of multiplicative perturbation.

\section{The chaotic neural network model}
The resonance phenomenon is also observed in an
excitatory-inhibitory neural pair,
with anti-symmetric, piecewise linear activation function.
This type of activation function has been chosen for ease of 
theoretical analysis. However, sigmoidal
activation functions also show similar resonance behavior.
\begin{figure}
\begin{center}
\leavevmode
\epsfig{file=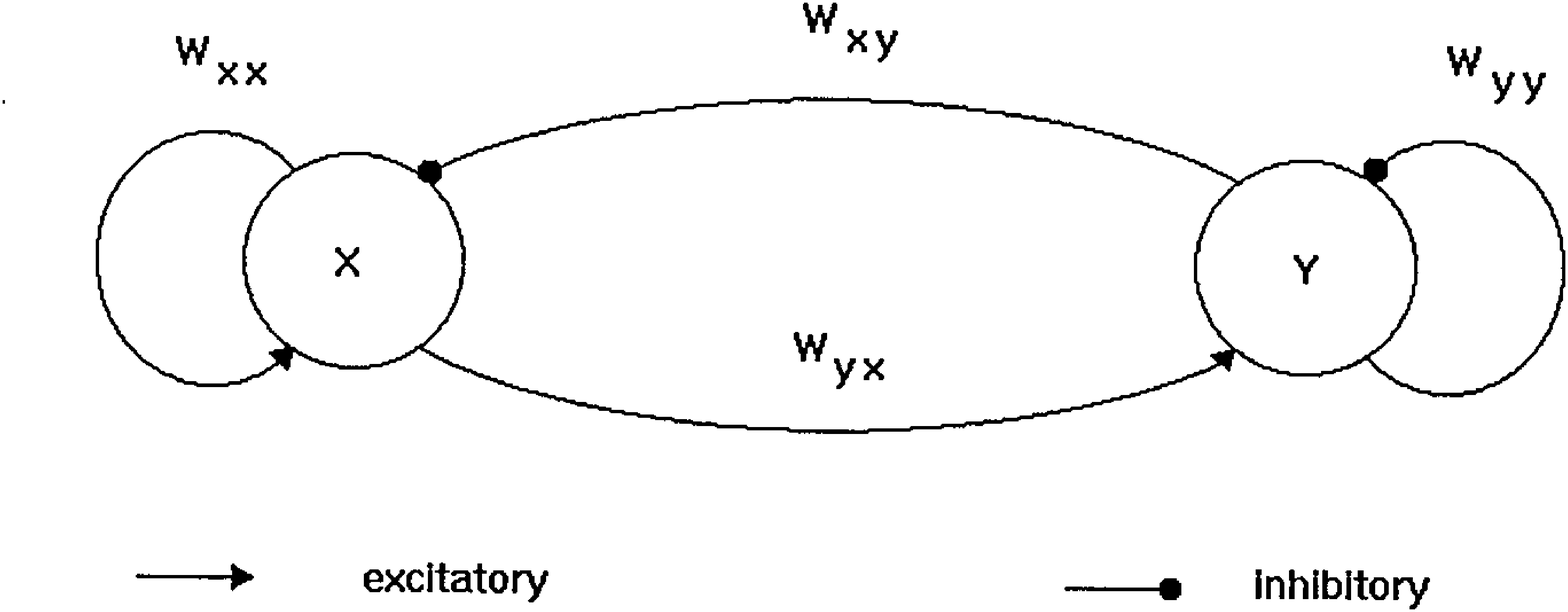, height=1.75in}
\end{center}
\caption{The pair of excitatory and inhibitory neurons used for
enhancing subthreshold signal.}
\label{exinpair}
\end{figure}
Fig. \ref{exinpair} shows a pair of coupled excitatory and inhibitory
neurons. The discrete-time dynamics of such simple neural networks are
found to exhibit a rich variety of behavior, including chaos \cite{Sin98b}.
If $x_n$ and $y_n$ ($x,y \in [-1,1]$) be the state of
the excitatory and inhibitory
elements at the $n$-th iteration, respectively, then the discrete
time-evolution equation of
the system is given by
$$ x_{n+1}= F_a (w_{xx} x_n - w_{xy} y_n + I_n),$$
$$ y_{n+1}= F_b (w_{yx} x_n - w_{yy} y_n + I_n),$$
where $w_{ij}$ is the connection weight from neuron $j$ to neuron $i$, and
$I$ is an external input. 
The activation function is of anti-symmetric, piecewise linear nature, viz.,
$ F_a (z) = -1, {\rm if}~ z < -1/a$, $F_a(z) = a z, {\rm if}~ -1/a 
\leq z \leq 1/a$, and $F_a(z) = 1, {\rm if}~ z > 1/a.$ 
Under the restriction $w_{xy}/w_{xx}=w_{yy}/w_{yx}=k$, the 2-dimensional
dynamics reduces to a simple 1-dimensional form. The relevant variable is
now the effective neural potential $z=x - ky$ ($z \in [-1,1]$),
whose dynamics is governed by the map
$$ z_{n+1} = {\mathcal F} (z_n) = F_a (z_n) - k F_b (z_n),$$
where $a,b$ are the suitably scaled transfer function parameters.
The design of the network ensures that the phase space [$-1+(kb/a),1-(kb/a)$]
is divided  
into two well-defined and segregated sub-intervals L:[$-1+(kb/a),0$] and
R:[$0,1-(kb/a)$].
The critical points of the map are at $z_c = \pm 1/a$. Note that,
if $\mathcal{F} (z_c)~(= 1 - \frac{kb}{a}) < \mathcal{F}^{-1}
(0)~(= \frac{1}{kb})$, the chaos is asymmetric, the trajectory being confined
to any one of the subintervals. Again, if $\mathcal{F} (z_c) > \frac{1}{b}$,
the trajectory will eventually converge to a superstable periodic cycle.
Therefore, for symmetric chaos, the following inequalities must be satisfied:
\begin{equation}
\frac{1}{kb} \leq 1 - \frac{kb}{a} \leq \frac{1}{b}.
\label{ineq}
\end{equation}
From the first inequality, we have $\frac{(kb)^2}{a} - kb + 1 \leq 0$,
and solving for $k$ in the limiting case of an equality, the $k$-value
at which the symmetry is just restored is obtained as
$k = \frac{a}{2b} (1 \pm \sqrt{1 - \frac{4}{a}})$. Note that,
real roots exist only for $a \geq 4$. Therefore,  
for $a < 4$, there is no dynamical connection
between the two sub-intervals and the trajectory, while chaotically wandering
over one of the sub intervals, cannot enter the other sub interval.
For $a > 4$, in a certain range of
$(b, k)$ values the system shows both symmetry-broken and symmetric
chaos, when the trajectory visits both sub intervals in turn. 
The curves in $(b/a, k)$-parameter space
forming a boundary between the symmetric and symmetry-broken chaotic
domains are given by $k = a (1 \pm \sqrt{1-(4/a)})/2b$.
The second inequality of (\ref{ineq}) gives, in the limiting case of
an equality, $k = \frac{a}{b} ( 1 - \frac{1}{b})$, which forms the boundary
between the regions showing symmetric chaos and superstable periodic cycles,
in the ($b/a, k$) parameter plane. 
The parameter space diagram in Fig. \ref{fig204iii} shows the various
dynamical regimes occurring for different values of $k$ and $b/a$,
at $a=6$.
\begin{figure}
\begin{center}
\leavevmode
\epsfig{file=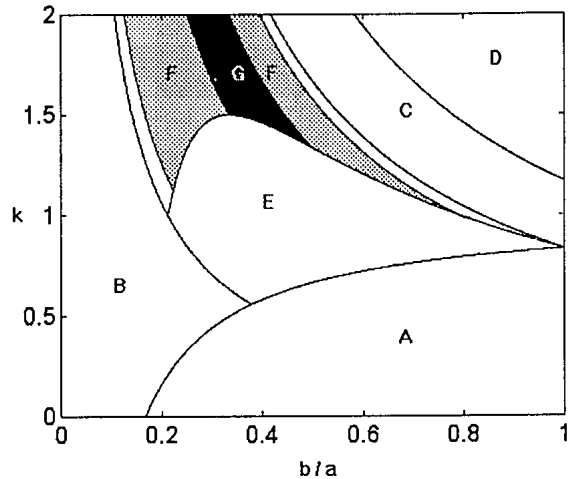, height=2.75in}
\end{center}
\caption{The ($b/a$) vs. $k$ parameter space at a = 6.0, for
neural pair dynamics governed by an anti-symmetric, piecewise
linear activation function. Region A: $z^* = 1 - k$ stable,
B: $z^* = 1/(1 + kb)$ stable, C: $z^* = 0$ stable, D: 2-period cycle
between $[(1-k),-(1-k)]$, E: superstable periodic cycles, F: two-band
symmetry-broken chaos, G: symmetric chaos. The two thin bands, between B and F,
and again, between F and C, indicate regions of single-band symmetry-broken
chaos.}
\label{fig204iii}
\end{figure}
For the simulations reported here,
$a=6$ and $b=3.42$, for which the system shows symmetric chaos
over a range of values of $k$.
 
\begin{figure}
\begin{center}
\leavevmode
\epsfig{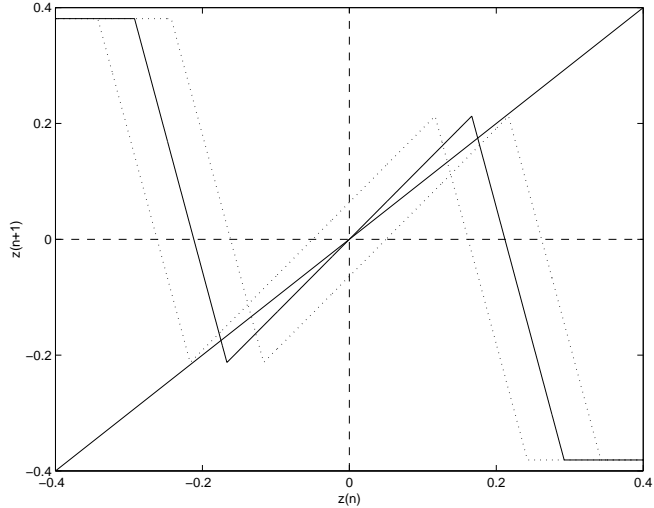}
\end{center}
\caption{The map representing the
dynamics of a neural pair for $a = 6.0, b = 3.42$
and $k = 1.3811$. The figure in solid lines represent the unperturbed
map ${\mathcal F}$, while the figures in dotted
lines indicate the maximum displacement
due to a periodic signal of peak amplitude, $\delta = 0.05$.}
\label{fig204iv}
\end{figure}

\begin{figure}
\begin{center}
\leavevmode
\epsfig{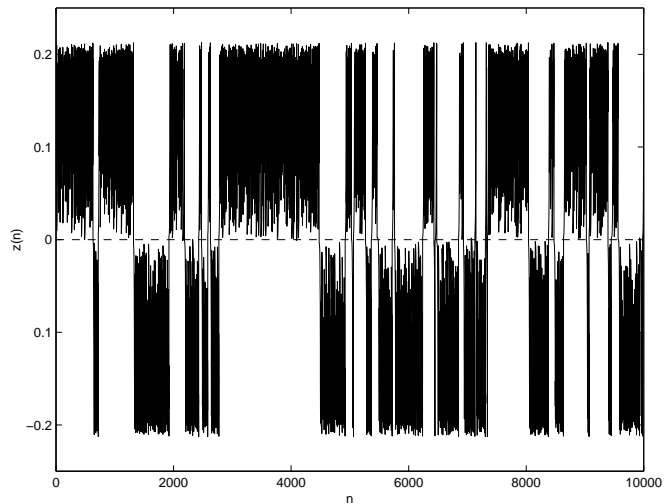}
\end{center}
\caption{The time-evolution of the sinusoidally perturbed neural pair
for $a = 6$, $b=3.42$, $k=1.3811$, $\omega = 1/400$ and $\delta = 0.0005$.
The broken line is the boundary between positive and negative
values of $z$.}
\label{figtime}
\end{figure}

The chaotic switching between the two sub-intervals occurs at random.
However the average time spent in any of the sub-intervals before a
switching event can be exactly calculated for the present model as
\begin{equation}
< n > =  \frac{1}{bk(1 - \frac{bk}{a})-1}.
\end{equation}
As a complete cycle would involve the system switching from one sub-interval
to the other and then switching back, the ``characteristic frequency'' of the
chaotic process is $\omega_c = \frac{1}{2 < n >}$. 
E.g., for the system to have a ``characteristic frequency'' of
$\omega = 1/400$ (say), the above relation
provides the value of $k \simeq 1.3811$ for $a=6, b= 3.42$.
The system being symmetric, there is no net drift
between L and R. However,
in the presence of an external signal of amplitude $\delta$, the symmetry
is broken. The net drift
rate, which measures the net fraction of phase space of one sub-interval
mapped to the other after one iteration, is given by
$v= \delta, {\rm if}~~\delta < {\delta}_c,$ and $v = 1-(kb/a)-
(1/bk),$ otherwise. The {\em critical signal strength},
\begin{equation}
{\delta}_c = 1 -(k^2 b^2 +a)/akb,
\end{equation}
is a limit above which the net drift rate no longer varies in phase
with the external signal. For the aforementioned system parameters $(a,b,k)$,
${\delta}_c \simeq 0.001$.
If the input to the system is a sinusoidal signal of amplitude $\delta <
{\delta}_c$ and frequency $\sim \omega_c$, we can expect the response to
the signal to be enhanced, as is borne out by numerical simulations.
The effect of a periodic input, $I_n = \delta \sin
(2 \pi \omega n)$, is to
translate the map describing the dynamics of the neural pair, to the
left and right, periodically. Fig. \ref{fig204iv} shows the unperturbed
map (solid lines) along with the maximum displacement to the left and
right (dotted lines) for $\delta=0.05$.
The resultant intermittent switching between the two sub intervals, L and R,
is shown in Fig. \ref{figtime} for $\omega = 1/400$ and $\delta = 0.0005 $.

\begin{figure}
\begin{center}
\leavevmode
\epsfig{file=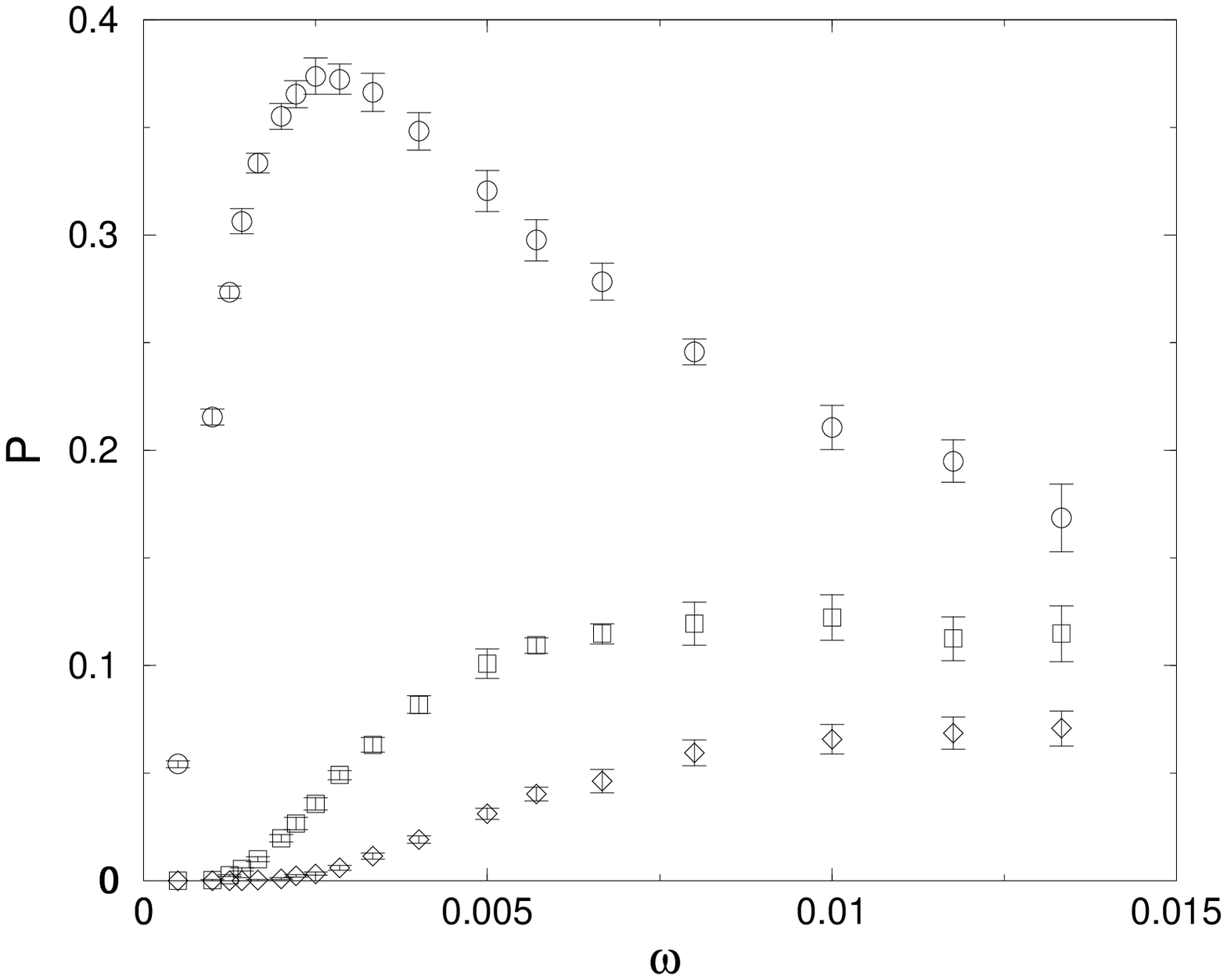, height=2.5in}
\end{center}
\caption{The peak strengths of the normalized residence time distribution,
$P_1$ (circles), $P_2$ (squares) and $P_3$ (diamonds),
for periodic stimulation of the excitatory-inhibitory neural pair
($a = 6$, $b = 3.42$ and $k = 1.3811$). The peak amplitude of the
periodic signal is $\delta$ = 0.0005.
$P_1$ shows a maximum at a signal frequency $\omega_c \simeq 1/400$.
Averaging is done over 18 different initial conditions, the error bars 
indicating the standard deviation.}
\label{fig205}
\end{figure}

As in the previous Section,
we verify the presence of resonance by looking at the
peaks of the residence time distribution, where the
strength of the $j$th peak is given by Eqn. (\ref{eq303}).
For maximum sensitivity, $\alpha$ is set as 0.25.
As seen in Fig. \ref{fig205}, the dependence of $P_j (j=1,2,3) $ on external
signal frequency, $\omega$, exhibits a
characteristic non-monotonic profile, indicating the occurrence of resonance
at $\omega \simeq \frac{1}{2 < n >}$. For the system parameters 
used in the simulation,
$< n > = 200$. The results clearly establish that the switching between states
is dominated by the sub-threshold periodic signal close to 
the resonant frequency.

\begin{figure}
\begin{center}
\leavevmode
\epsfig{file=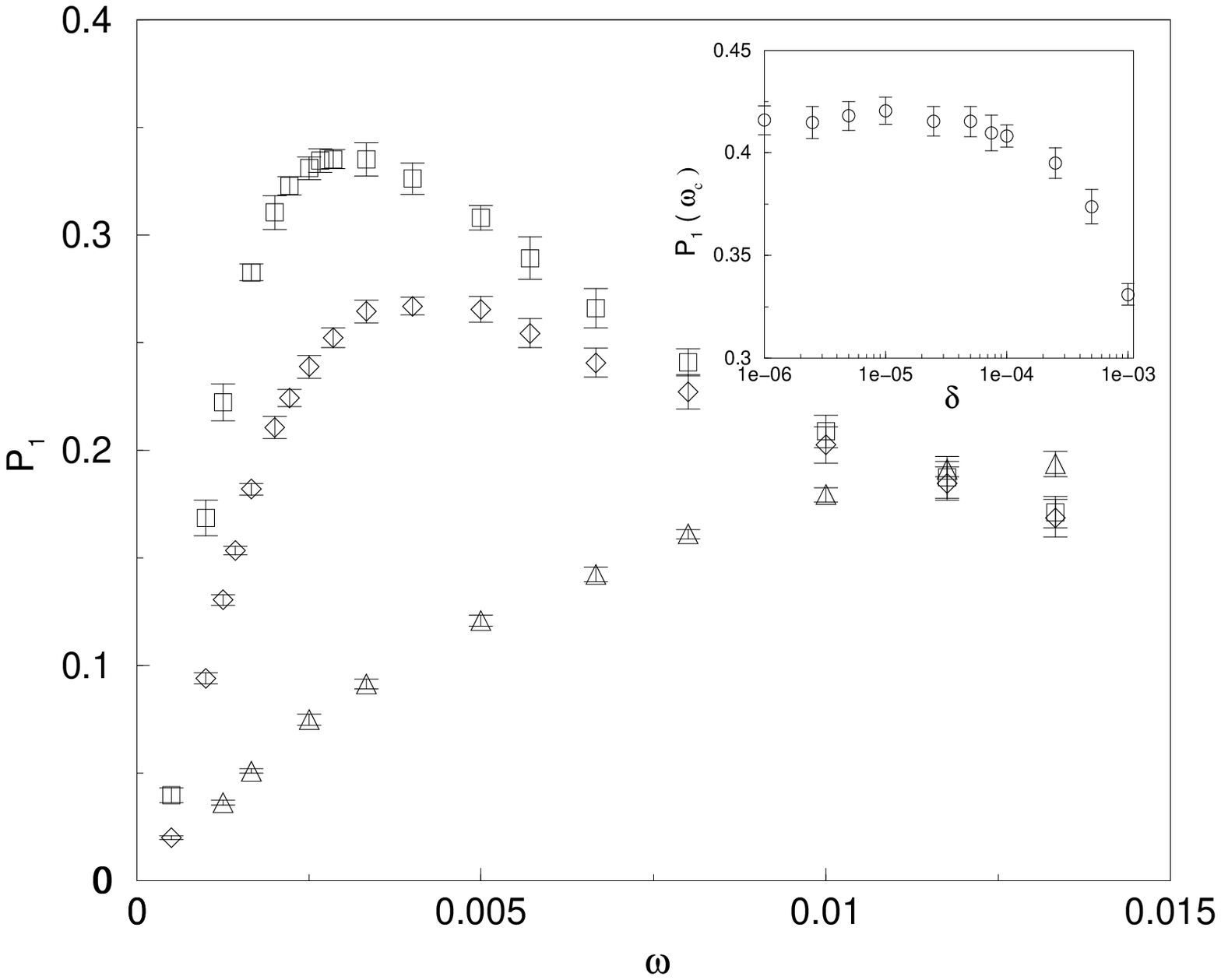, height=2.5in}
\end{center}
\caption{The strength of the first peak ($P_1$) in the normalized
residence time distribution for periodic stimulation at $\delta=0.001$
(squares), $=0.0025$ (diamonds) and $=0.01$ (triangles). The inset shows
$P_1$ at ${\omega}_c=1/400$ against $\delta$. System parameters same as
in Fig. \ref{fig205}}.
\label{fig205a}
\end{figure}
The variation of $P_1$ with $\omega$ for different values of signal
amplitude, $\delta$, was also studied. For $\delta > {\delta}_c \simeq 0.001$,
the variation in the drift rate no longer matches the signal, and
the maximum response is found to shift to higher frequency values 
(Fig. \ref{fig205a}): e.g., at $\delta = 0.01$, the maximum response in
$P_1$ occurs at $\omega \simeq 0.03 $. For $\delta < \delta_c$,
the magnitude of $P_1$ at the
resonance frequency, ${\omega}_c$, has a non-monotonic nature
(Fig. \ref{fig205a}, inset). For the system parameters mentioned here,
the maximum response occurs at $\delta \sim 10^{-5}$.   

These results assume significance in light of the work done on detecting
SR in the biological world. In neuronal systems, a non-zero SNR is found
even when the external noise is set to zero \cite{Wie95}.
This is believed to be due to the existence of ``internal noise". This
phenomenon has been examined through neural network modeling, e.g., in
\cite{Wan97}, where the main source of such ``noise" is the effect of
activities of adjacent neurons. The total synaptic input to a neuron,
due to its excitatory and inhibitory interactions with other neurons,
turns out to be aperiodic
and noise-like. The neural network model employed in the present work is,
however, the
simplest system to date, which uses its aperiodic activity
to show SR-like behavior. There is also a possible connection of
such `resonance' to
the occurrence of epilepsy, whose principal feature is the synchronization
of activity in neurons.
\section{Discussion}
Low-dimensional discrete-time dynamical systems
are amenable to several analytical
techniques and hence can be well-understood compared to other systems.
The examination of resonance phenomena in this scenario was for ease of
numerical and theoretical analysis. However, it is reasonable to assume that
similar behavior occurs in higher-dimensional chaotic system, described by
both maps and differential equations. In fact, SR has been reported
for spatially extended systems (spatiotemporal SR) \cite{Lin95},
e.g., in coupled map lattices \cite{Gad97}.
A possible area of future
work is the demonstration of phenomena analogous to spatiotemporal SR
with a network of coupled excitatory-inhibitory neural pairs.

The above results indicate that deterministic chaos can play a constructive
role in the processing of sub-threshold signals. 
It has been proposed that the
sensory apparatus of several creatures use SR to enhance their sensitivity
to weak external stimulus, e.g., the approach of a predator.
Experimental study involving
crayfish mechanoreceptor cells  have provided evidence of
SR in the presence of external noise
and periodic stimuli \cite{Dou93}. 
The above study indicates that external noise is not necessary for such
amplification as chaos in neural networks can enhance weak signals.
The evidence of chaotic
activity in neural processes of the crayfish \cite{Pei96} suggests
that nonlinear
resonance (as reported here) due to inherent chaos might
be playing an active role in such systems. 
As chaotic behavior is extremely common in a recurrent network
of excitatory and inhibitory neurons, such a scenario is not entirely
unlikely to have occurred
in the biological world. This can however be confirmed only by further
biological studies and detailed modeling of the phenomena.

{\bf Acknowledgements:} I would like to thank Prof. B. K. Chakrabarti
for many helpful discussions and Abhishek Dhar for a careful reading
of the manuscript. JNCASR is acknowledged for
financial support.

\end{document}